# Deep reinforcement learning for efficient exploration of combinatorial structural design spaces

Chloe S.H. Hong*, Keith J. Lee, Caitlin T. Mueller

*Massachusetts Institute of Technology
77 Massachusetts Ave, Cambridge, MA 02139
cshhong@mit.edu

**Abstract**
This paper proposes a reinforcement learning framework for performance-driven structural design that combines bottom-up design generation with learned strategies to efficiently search large combinatorial design spaces. Motivated by the limitations of conventional top-down approaches such as optimization, the framework instead models structures as compositions of predefined elements, aligning form finding with practical constraints like constructability and component reuse. With the formulation of the design task as a sequential decision-making problem and a human learning inspired training algorithm, the method adapts reinforcement learning for structural design. The framework is demonstrated by designing steel braced truss frame cantilever structures, where trained policies consistently generate distinct, high-performing designs that display structural performance and material efficiency with the use of structural strategies that align with known engineering principles. Further analysis shows that the agent efficiently narrows its search to promising regions of the design space, revealing transferable structural knowledge.

## 1. Introduction

Computational methods have advanced performance-driven structural design, particularly by using form to enhance structural efficiency and reduce material use. Many of the existing approaches rely on top-down design generation via parameterization coupled with gradient-based optimization. However, this approach is limiting both in terms of the narrow expressivity and diversity that can be encoded in a parametric design space, and in terms of the disconnect between global form representation and eventual materialization into component parts.

The motivation of this work is to explore alternative computational performance-driven design frameworks where designs are premised as compositions of predefined elements, and high performing designs are found by exploration of the large design space consisting of all combinatorial possibilities. Such an approach enables the form-finding process to be inherently aligned with real-world construction constraints, avoiding the need to retrospectively align the building with the form and compromise critical factors such as constructability and reuse as seen with predominant approaches. Achieving this, however, requires a shift from top-down design generation to a bottom-up process, and from purely gradient-based optimization to methods that balance exploration and exploitation. To this end, the paper proposes a reinforcement learning framework for performance driven structural design that combines bottom-up design generation with learned heuristics to efficiently guide the search.

## 2. Related work

### 2.1. Performance driven design with bottom-up design generation

Bottom-up design generation methods involve incremental application of rules to create structural configurations. The key strength compared to top-down parametric approaches, such as shape optimization which fixes topology and varies dimensions, is that bottom-up methods define form through growth, aggregation, or grammar-based processes enabling the exploration of combinatorially rich design spaces. Shape grammars [1], for instance, are capable of creating infinite design variants with recursive shape-driven design rules, as shown with Mueller's work on generating spanning structures across typologies [2] and Lee's work [3] to generate a range of equilibrium forms from force-path grammars. Another key advantage is that it possible to enforce local constraints such as constructability or partial-state stability at each step of the design generation, effectively reducing the search space to feasible, sensible designs. For instance, Lee [4] presents a growth-based algorithm that enforces partial-state stiffness and constructability to inform the branch-and-bound search for truss design and Wasp [5] proposes design from an aggregation of components limited by local connectivity rules. While bottom-up methods enable the generation of diverse and feasible designs, efficiently guiding this process toward high-performing solutions remains a core challenge. The vast combinatorial space driven by variations in element types, compositions, and dimensions makes the search exponentially complex, with multiple local optima further complicating convergence. An effective approach must not only broadly explore this space but also prioritize promising design trajectories.

A previously attempted strategy for performance-driven bottom-up design involves coupling generative rules with stochastic optimization methods such as genetic algorithms. The stochasticity in design generation along with mutation and crossover operations allows exploration of the large design space, while selection mechanisms gradually guide the process toward high-performing designs. In structural design, [2], [6] demonstrate this by employing subjective human input and objective structural performance metrics to steer the search for grammatical truss design. However, the inconsistent results show that selection alone is an unreliable optimization strategy, often requiring highly curated grammars to achieve reliable convergence. Even when successful, the stochastic nature of operations makes it difficult to interpret, reproduce, or systematically refine the relationship between design outcomes and the rules that generated them.

### 2.2. Design as a sequential decision-making task solved with reinforcement learning

The need to tightly couple the generated design with the decision-making process motivates the use of reinforcement learning (RL), which optimizes the final outcome by the sequence of actions that lead to it. Originally developed in operations research and integrated with neural networks, Deep RL has shown success in tasks in robotics, strategy board games, and video games involving long horizon decision-making in complex dynamic systems. These task characteristics align with performance driven bottom-up design where the goal is to find an unobvious sequence of rules to generate optimal results.

The application of reinforcement learning to structural design remains limited and requires task-specific adaption of representations, algorithms, and training algorithms. Performance-driven structural design poses unique challenges, mainly due to the highly nonlinear landscape that solutions lie in. Not only is the combinatorial design space large, but multiple high performing designs are sparsely distributed within a sea of infeasible or poorly performing designs. A limited number of studies illustrate RL's potential in this context. Zhao *et al.* [7] demonstrate that the algorithmic frame of RL to train a neural network based heuristic for guiding rule selection, as used in strategy board game-playing systems is transferable to performance-driven robot morphology generation. Hayashi *et al.* [8] uses RL to obtain sparse truss topologies minimizes the total structural volume under stress constraints by iteratively removing elements from a ground structure while representing partial designs as graphs. Similarly Brown *et al.* [9] employs Double Q-learning to replicate topology optimization on a binary 2D grid, introducing progressive resolution refinement to address the computational bottleneck of frequent finite element analysis evaluations. Bapst *et al.* [10] applies RL to physical construction tasks involving the assembly of rigid bodies from a discrete inventory to satisfy functional objectives such as spanning, coverage, or silhouette formation under gravity. More recently, Sørensen *et al.* [11] applies RL for the inventory-constrained design of planar roof trusses where the emphasis is on creating designs fit to trajectories using a set of inventory with deflection constraints. Related to graphic statics, Tam et al. [12] employ Graph Neural Networks and RL to inverse design shell structures and Rajasekar et al. [13] propose using RL to guide grammatical design. Collectively, these works demonstrate RL's potential for tackling combinatorial structure generation and provide insight into tailored representations and algorithms that enable reasoning about geometry and spatial behavior, capabilities that are essential for structural design.

## 2.3. Research gap and opportunities

This paper presents a reinforcement learning framework to address the absence of integrated, performance-driven bottom-up design frameworks that can guide the step-by-step generation of structural designs composed of varying discrete, predefined elements. In current practice, high performing designs with discrete elements require a costly combinatorial search, often constrained to a limited set of predefined geometries. This raises key questions: Can we computationally learn strategies i.e., sequences of rule applications, that lead to high-performing designs? Is it possible to identify promising rules at each design step based on feedback from prior experience? Crucially, this work leverages the unique strengths of RL, a framework for optimizing sequential decision-making by learning and reusing strategies through guided trial and error, to enable efficient exploration and adaptation across diverse design scenarios. In doing so, RL is positioned not simply as a tool for solution matching, but as a framework for exploratory problem-solving in complex, combinatorial structural design tasks.

## 3. Methodology

### 3.1 Representation of structural design as a Markov decision process

*3.1.1. Modular truss frame structure*
The case study considers designing cantilever structures assembled from modular truss frames. Each frame is composed of hollow circular steel tube elements with a Young's modulus of 200 GPa, shear modulus of 80 GPa, and a yield strength of 350 MPa. The sectional area is calculated as $A = \pi (R^2 - R_{inner}^2)$ and moment of inertia with $I = (\pi / 4)(R^4 - R_{inner}^4)$ with $R_{inner} = (1 - \alpha) R$ where $R$ and $R_{inner}$ denote the outer and inner radii of the steel tube and $\alpha \in (0, 1)$ is the wall-thickness ratio. For this study, two frame types composed of tubes of different sections are employed as shown with Table *1*. In each episode, the total inventory of light and medium frames is fixed at the start.

Table 1: Parameters for section geometry of two frame types used in the case study.

| Frame type | Outer radius R (m) | Thickness ratio α | Self-load (kN) * applied per node |
|---|---|---|---|
| Light | 0.10 | 0.10 | 4 |
| Medium | 0.20 | 0.10 | 6 |

*3.1.2. Markov decision process*
This work builds on existing frames of reinforcement learning where the goal is to find the optimal policy of a Markov decision process (MDP). A MDP is characterized as the tuple $\mathcal{M} = (\mathcal{S}, \mathcal{A}, P, R)$, where $\mathcal{S}$ is the set of states, $\mathcal{A}$ is the set of actions, P: $\mathcal{S} \times \mathcal{A} \times \mathcal{S} \rightarrow [0,1]$ is the transition kernel, with P(s' | s, a) = P($s_{t+1}$ = s' | $s_t$ = s, $a_t$ = a), and R: $\mathcal{S} \times \mathcal{A} \rightarrow \mathbb{R}$ is the (possibly stochastic) reward function,

with $r_t = R(s_t, a_t)$. At each step t, the agent observes a state $s_t \in \mathcal{S}$, selects an action $a_t \in \mathcal{A}$, receives reward $r_t = R(s_t, a_t)$, and transitions to $s_{t+1} \sim P(\cdot \mid s_t, a_t)$. The optimal policy $\pi^*$ is a probabilistic mapping of states to actions such that when followed would return the maximal expected cumulative rewards.

### 3.1.3. State: frame grid

Each state $s_t$ is represented as an integer grid vector $D_t$, which encodes information about the boundary condition (locations of supports and external loads, and external load magnitudes), the current design configuration, and optionally the remaining inventory, as shown with Figure *1*. Table *2* presents a mapping of frame types with two frame types, denoted as light and medium where each cell $(i, j)$ of $D_t$ takes one of these integer values.

Table 2: Example of index mapping of frame types for state grid vector representation used in the case study.

| Value | Description |
| --- | --- |
| -1 | External load |
| 0 | Unoccupied |
| 1 | Support frame |
| 2 | Free frame (light) |
| 3 | Free frame (medium) |
| 4 | (optional) Inventory slot: free frame (light) |
| 5 | (optional) Inventory slot: free frame (medium) |

The support frame is differentiated as it is defined to have fixed nodal connections, as opposed to other frames that are free. Although not physically a frame, the location of external forces is indicated as a cell value of –1. While two types, light and medium frames, are used in this study, any variety of modules can be employed. The frame grid is defined to include information about the current inventory, even if some elements are not yet part of the design at the current state, as this information is useful when planning the sequence of frame addition ahead of time.

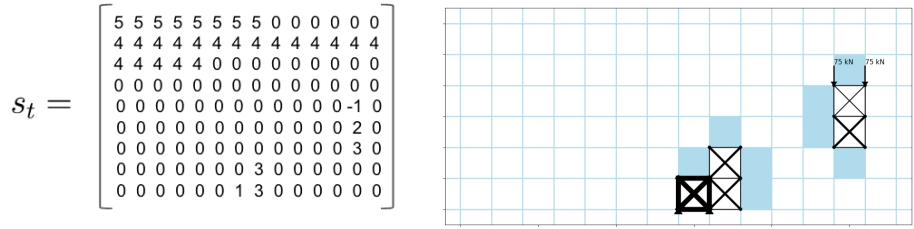

Figure 1: Example of a state grid vector corresponding to a partial design state. The state grid vector (left) concatenates the current inventory stock (first 3 rows) and the occupancy grid of the partial design (last 6 rows). In the example partial design state, 3/20 light frames, and 1/10 medium frames are used. In the visualization of the partial design state (right), sections of the frames are differentiated by line thicknesses and light blue cells indicate the allowed locations for next frame placement.

### 3.1.4. Action: placement of truss frames

An action at step $t$ is specified by the tuple

$$a_t = (e_t, k_t, i_t, j_t)$$

where $e_t \in \{0, 1\}$ is a termination flag ($e_t = 1$ ends the episode), $k_t \in \{2, 3\}$ denotes the free-frame type index (2 = light, 3 = medium), and $(i_t, j_t)$ identifies the grid cell in which the frame is placed. For instance, in a 6 × 14 design region of the grid, $i_t \in \{1, ..., 6\}$ and $j_t \in \{1, ..., 14\}$.

*3.1.5. Transition probabilities: action masking*

Transition probabilities are defined only for actions that maintain connectivity and respect inventory limits. Naïvely allowing any frame type to be placed in any grid cell would result in disconnected designs and exceed available stock. Instead, for each state *s*, the transition probability is defined as $P(s' \mid s, a)$ which is 1 if all feasibility conditions hold, and 0 otherwise. The feasibility conditions for an action *a = (e, k, i, j)* are:

Condition 1) Cell (*i, j*) lies adjacent to the cells occupied by the existing design.
Condition 2) Frame type *k* has at least one unit remaining in inventory.
Condition 3) Termination flag *e* = 1 is permitted only if all supports and targets are connected.

This form of "hard" masking ensures that at every step, the agent samples only from feasible actions; those that maintain structural connectivity and remain within the bounds of the available inventory.

*3.1.6. Reward scheme*

At each step of the MDP a reward is given. In this study, there are *interim rewards* given at non-terminal states and *terminal rewards*. The interim reward encourages the agent to build towards connecting the support and target locations. The interim reward is defined as $r_{t\_int} = 0.0025 \times f\_conn(s_{t+1})$ where $f\_conn(s) \in [0, 1]$ is the fraction of target nodes connected to the support in state *s*. While this interim reward provides a helpful training signal in a long-horizon problem, the small coefficient ensures its magnitude remains negligible compared to the terminal reward.

At episode termination, which is permitted only when a feasible design connects the support and target within the available frame inventory, finite element analysis (FEA) is performed to return information of nodal deflection and elemental stiffness, as shown with Figure 2. If the design remains incomplete and the inventory has been exhausted, the episode is truncated, and the reward is set to zero. Otherwise, the terminal reward is a weighted sum of elements stated in Table 3. This reward formulation was developed through iterative tuning to discourage adversarial behaviors, such as generating monolithic structures or designs that fail to meet stiffness requirements. Note that the inventory-usage penalty component is capped, the deflection penalty is either {0, −1} while the element failure penalty scales linearly with the number of failed elements, i.e. those where the stress limits are exceeded. The weighting reflects a prioritized design objective: first eliminating structural failures, then preventing excessive deflection, and finally minimizing material usage.

Table 3: Description of terms composing the terminal reward.

| Reward Component | Equation | Description |
| --- | --- | --- |
| Target completion | $+N_{target}$ | Rewards connecting supports to target locations |
| Inventory usage penalty | $-N_{used}/N_{inventory}$ | Penalizes proportion of inventory used |
| Deflection penalty | $-1\{\delta_{max} \geq \delta_{allowable}\}$ where $\delta_{max}$ is the maximum nodal displacement of the design and $\delta_{allowable}$ is the allowable displacement which is calculated as $1/120$ of the cantilever length | Penalizes designs that exceed deflection limit |
| Element failure penalty | $-|\mathcal{F}|$ where $\mathcal{F}$ is the set of elements that exceed their allowable axial stress | Penalizes the number of failed elements |

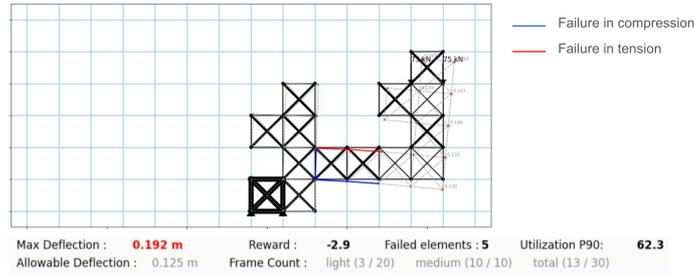

Figure 2: At termination of episode, FEA is performed to return information of nodal max deflection, number of failed (those that exceed their allowable stress limit) elements, 90 percentile of elemental utilization, and inventory usage.

## 3.2 Two phase learning

This paper introduces a two-phase training process, with a novel task division scheme inspired by human learning: the reasoning task is split and the agent reuses learned strategies in the training that enables efficient search for high-performing designs. In the first phase, a base policy is trained on a simplified yet adaptive task: generating a feasible design that connects the support and target load placed at random locations within the grid, under self-load conditions and using a single frame type. In the second phase, this policy is fine-tuned to solve the more constrained task of connecting a fixed target position under external loading, with the option to employ varying frame types. By dividing the complex reasoning task into distinct phases, the agent's search is effectively guided toward promising regions of the design space, without being confined to a single local optimum, allowing efficient optimization across a large combinatorial space. Compared to the alternative of training a separate policy from scratch for each boundary condition, this phased approach results in performant designs for generalized conditions more reliably and efficiently.

The training algorithm in both phases is carried out using Proximal Policy Optimization (PPO) [14] with an actor-critic architecture where the policy (actor) and value (critic) networks share a convolutional neural network (CNN) backbone. The shared encoder consists of three convolutional layers with channel widths of 64, 128, and 128, respectively, each followed by layer normalization and ReLU activation followed by a fully connected layer of width 512, from which separate linear heads branch off for the actor and critic. These heads are conditioned on design scenario-specific constraints, including the target location and available inventory.

**Algorithm 1** Training Base Networks
Randomized Target, Self-Load, Single Frame Type
**Output:** Trained policy $\pi_\theta$ and value network $V_\phi$
**Initialize** $\pi_\theta$ and $V_\phi$ from scratch
**for** $j \leftarrow 1$ **to** $N$ **do**
    ▷ **Rollout**
    Randomize target position and initialize start state
    Initialize inventory with one frame type
    Take $n_{\text{rand}}$ random steps at initialization
    Follow $\varepsilon$-greedy policy until termination or truncation
    Compute reward via FEA or assign 0 if truncated

    ▷ **Update**
    Train $\pi_\theta$ and $V_\phi$ on collected transitions

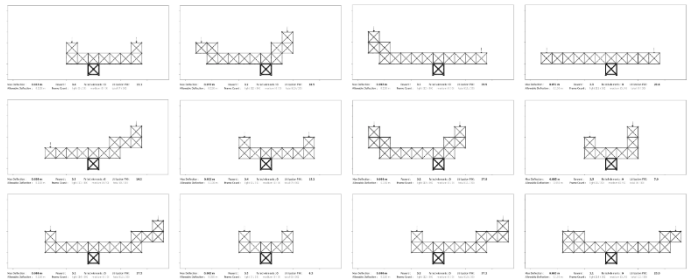

```
Algorithm 2 Transfer Learning
Fixed Target Location, External Load, Mixed Frame Type
Output: Adapted policy π_θ and value network V_φ
Initialize π_θ ← π_base, V_φ ← V_base
for j ← 1 to N do
    ▷ Rollout
    Fix target at (h, l) with external load P
    Initialize inventory with multiple frame types
    Follow ε-greedy policy until termination or truncation
    Compute reward via FEA or assign 0 if truncated

    ▷ Update
    Train π_θ and V_φ on collected transitions
```

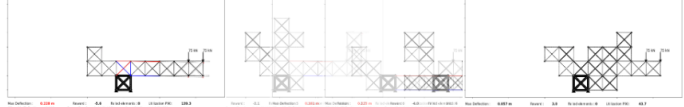

Figure 3: Phase 1 trains policies to create feasible designs under self-weight for various target locations and Phase 2 fine tunes that policy for individual design scenarios to withstand external loads.

As shown in Figure 3, in the first phase, a base policy network $\pi\_\theta(a \mid s)$ and a value network $V\_\phi(s)$ are optimized over $N$ iterations, with $n_{rollout}$ steps for each iteration. Episodes are initialized with a fixed support location and random target positions within the grid. Each episode begins with $n_{rand}$ steps of random actions to promote design diversity and reduce the likelihood of early convergence to a local optimum. Following this, actions are selected using an ε-greedy strategy. Once transitions are collected, the networks are trained for $n_{epochs}$ epochs using minibatches of size $M$ sampled from the transition buffer. The policy network is updated using a clipped surrogate objective, which constrains policy updates by bounding the probability ratio between new and old policies. The value network is trained using a squared error loss, following the standard PPO framework. In the second phase, the weights of the base policy network and value network trained in Phase 1, are used to initialize fine-tuning. The overall training structure remains similar to Phase 1, with the following modification: within the rollout, each episode is initialized with a fixed support and a specified target location defined by a height and length pair $(h, l)$, along with an external load of magnitude $P$ and an inventory containing multiple frame types. The remainder of the rollout and training steps is identical to those in Phase 1.

## 4. Results

### 4.1 Implementation

Experiments were conducted locally on a MacBook Pro with an Apple M1 Pro chip (8-core CPU with 6 performance and 2 efficiency cores), a 14-core GPU, and 16GB of unified memory with 200GB/s memory bandwidth. The training of the base networks consists of 100,000 rollout steps, with approximately 6400 episodes, taking 200 minutes in total. The following transfer learning consists of additional 50,000 rollout steps with 3000-4000 episodes taking 30-40 minutes. Within each episode, at each step, an action is inferred from the current neural network policy with the episode lengths vary from 12-30 steps. In the case that the episode terminates with a feasible design, the FEA conducted with ASAP [15] takes approximately 0.4-0.5 seconds, being the most time-consuming part the run of each episode.

### 4.2 Case study

As a demonstration of the proposed framework, a base policy is trained for 100,000 training steps and fine-tuned for 50,000 training steps individually on 12 different design scenarios for cantilevers with varying target locations and external loads to examine how the trained policy effectively finds distinct high performing designs, that a stochastic search within a reasonably constrained design space, represented by the baseline method, is unable to find. In this context, high performance of a design is defined as meeting the stiffness required by the boundary conditions while minimizing the use of material. Generated results are compared to those from a baseline which is devised as a reasonable proxy of what a human designer would do naively. As shown with Figure 4, The baseline design generation logic involves stochastically expanding and permuting frame types from a randomized Manhattan path connecting the support and target locations.

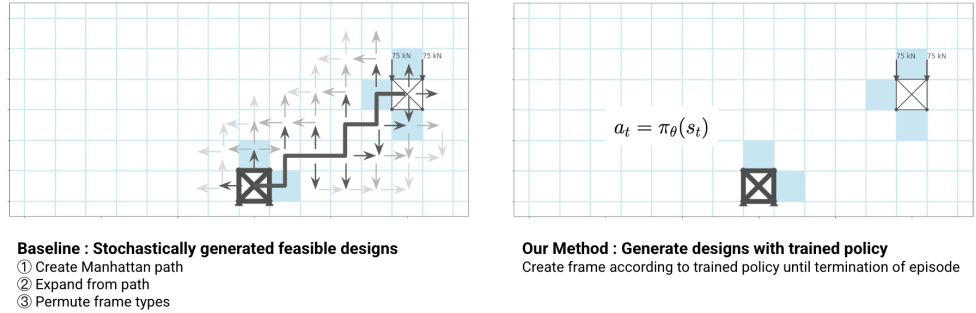

Figure 4: Design generation logic for baseline that creates randomized feasible designs and the proposed method that sequentially generates designs by running inference with the trained policies.

By default, all 12 design scenarios have two target locations, each on the left and right side of the support which are varied by their height and length from the support and applied external load in magnitude 100kN, 150kN, 200kN. While there is at least one external load for each scenario, targets that do not have any external load can be interpreted as positions for the structure to reach, as specified by a designer for aesthetic or other functional reasons. This also allows the method to generate diverse results for a given boundary condition. The varying external loads and positions impose different levels of task difficulty, which can be retrospectively characterized by the average number of failed elements in baseline-generated designs. Design scenarios that more frequently have elements that exceed their stress limit are assumed to be more difficult.

A selection of inferred designs for different design scenario is shown in Figure 5. The fact that the generated designs vary vastly between different scenarios illustrates that even when the conditions change slightly, an effective design can be drastically different, and the proposed method is capable finding tailored high performing design for each specific boundary conditions. Visibly, designs from the random baseline show large deformations in Figure 5 with many elements that exceed the allowable stress capacity and frequently use more material, while the instances from the proposed method have minimal failed elements and do not use material excessively (Table 4, Figure 6). Significantly, while have varying forms per scenario, solutions from the proposed method exhibit common strategies of material allocation as placing stronger frames towards the base, where shear and moment demands are maximum and creating a tapering shape either on the top or bottom side leading to the cantilever. Design scenarios 4 and 12 also introduce voids where structural material is less valuable.

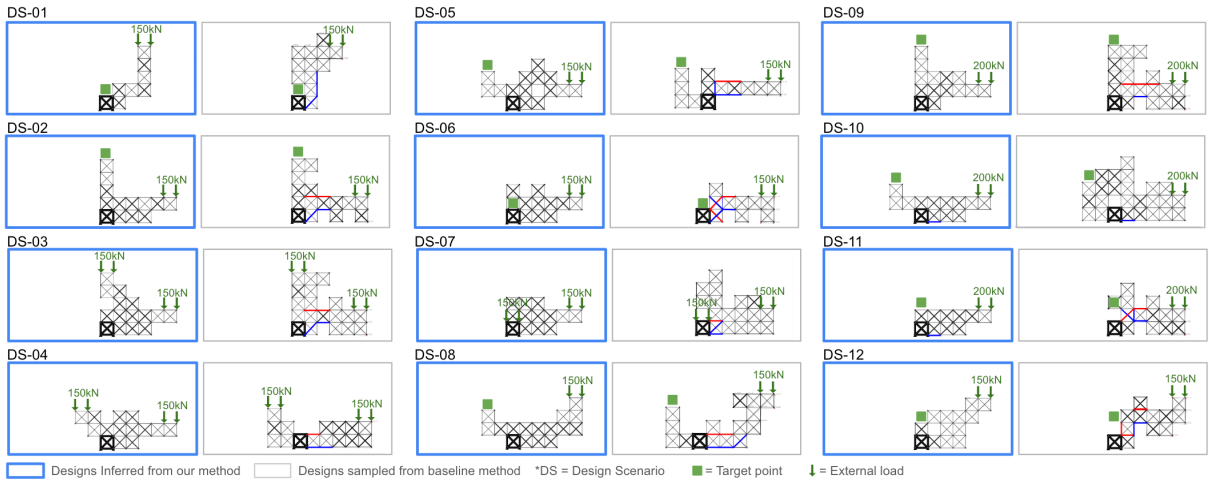

Figure 5: Selective designs inferred from trained policy (left outlined in blue) and baseline method (right) for design scenarios in order of task difficulty. Design scenarios all form cantilevers but have varying target and external load locations and magnitudes. Inferred designs use fewer frames and rarely have failed elements whereas the baseline designs use excessive number of frames and have multiple failed elements.

Quantitative comparisons between designs generated by our trained policies and the baseline reveal consistent improvements in both structural performance and material efficiency (Table 4, Figure 6). Most notably, the average number of failed elements per design is significantly reduced by 3.5 with improvements ranging from 1.2 to 5.5 fewer failed elements across all boundary conditions, demonstrating a universal reduction in structural failure. In terms of material use, our method yields designs with consistently fewer frame elements, achieving an average reduction of 5.4 frames per design. This suggests that the trained policy not only improves feasibility but also promotes more economical structural configurations. Finally, the 90th percentile of elemental utilization, a measure of how efficiently each element's capacity is used, shows a notable average improvement of 12.5% over the baseline, highlighting the ability of our method to generate structurally optimized and efficient solutions.

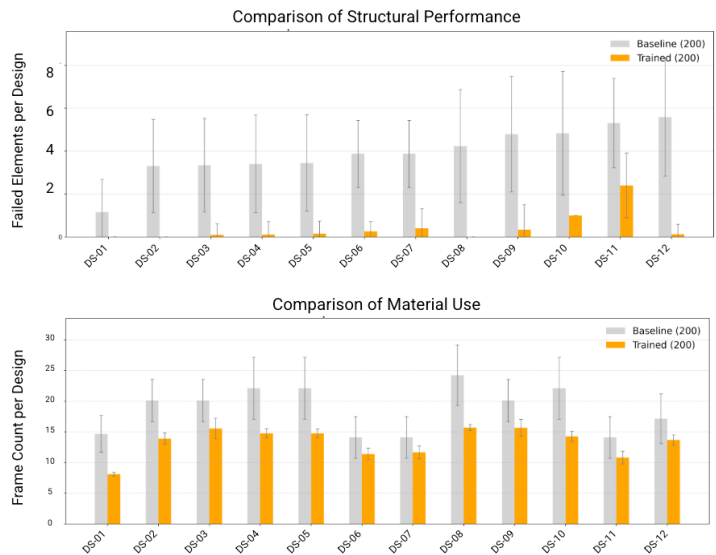

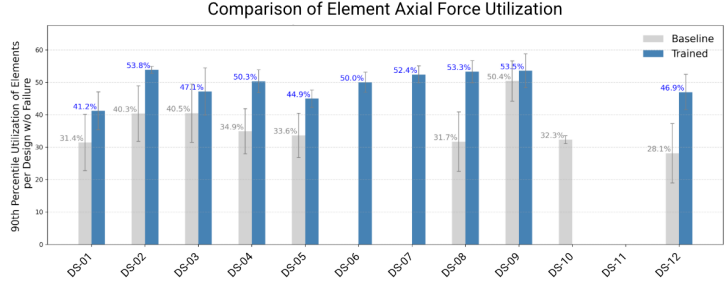

| Metric | Avg Δ | Max Δ | Min Δ |
|---|---|---|---|
| Utilization (90th percentile) | **+12.53%** | +21.58% | +3.16% |
| Avg. Failed Elements | **-3.52** | -5.46 | -1.15 |
| Avg. Frame Count | **-5.40** | -8.53 | -2.45 |
| % Solutions w/o Failed Elements | +68.39% | +91.15% | -2.09% |
| % Solutions within Allowable Deflection | +62.97% | +83.29% | +48.45% |

* Allowable deflection is $\frac{1}{120}$ of cantilever length

Figure 6, Table 4 : Comparison of the structural performance and material use of designs inferred from trained policy with baseline method in terms of average number of failed elements (top), average frame count (middle), and 90 percentile elemental utilization (bottom).

### 4.3 Efficient search of the design space

Visualization of the design space with instances encountered during the training of the agent verifies that RL's distinct mechanism of policy learning efficiently searches the large combinatorial space. The design space is visualized in a 2D space where the high dimensional vector grid representation of each design is reduced to $\mathbb{R}^2$ with the nonlinear dimension reduction technique Uniform Manifold Approximation and Projection (UMAP) [16]. As retrieving the entire design space with all possible designs is intractable, a reasonable approximation of its bounds can be made by sampling a large number, 500, of designs from the baseline method, which is devised to randomly generate feasible designs. The approximated design space is overlayed with an equal number of designs inferred across policies during the training of the agent each at 7500 step intervals within the 50,000 steps of training of Phase 2. These designs effectively represent the designs that are encountered during the trial-and-error training process of the policy and equivalently the search of the proposed method within the design space. All designs shown are for a single design scenario, DS-05 (defined in Figure 5).

Observation of search patterns throughout the course of the training of the agent in the second phase, clearly show strategic and efficient search. Figure 7 shows that initially the span of the search, defined as the minimally bounding circle of all instances, encompassing the bounds of the entire design space but soon the span is narrowed down to selective regions while identifying targeted design instances that eventually all meet the stiffness requirements.

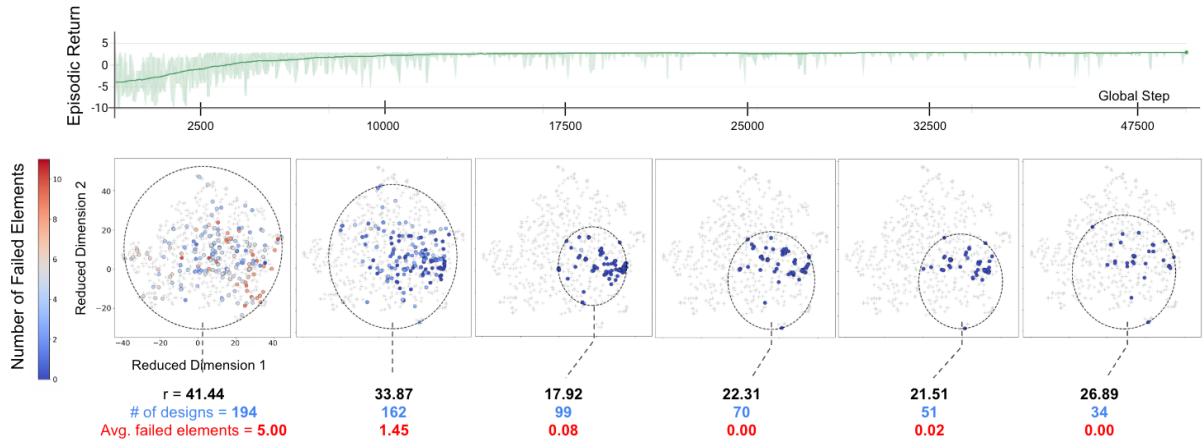

Figure 7: Visualization of explored designs within the design space during the training of Phase 2. The breadth of the search, indicated with the radius of the minimal bounding circle, initially encompasses the design space, then rapidly reduces to include selective designs that meet the required stiffness.

Figure 8 shows that designs inferred from the final trained policy represent several high-performing solutions that seem to lie along the Pareto front of the reward function. These instances clearly exhibit effective structural strategies, such as overall tapering geometries and the placement of stronger frames near the base, which contribute to the required stiffness of the cantilever while minimizing material use. The learned structural strategies are evident when compared to baseline designs, which either fail due to poorly placed material or require excessive material to meet performance criteria.

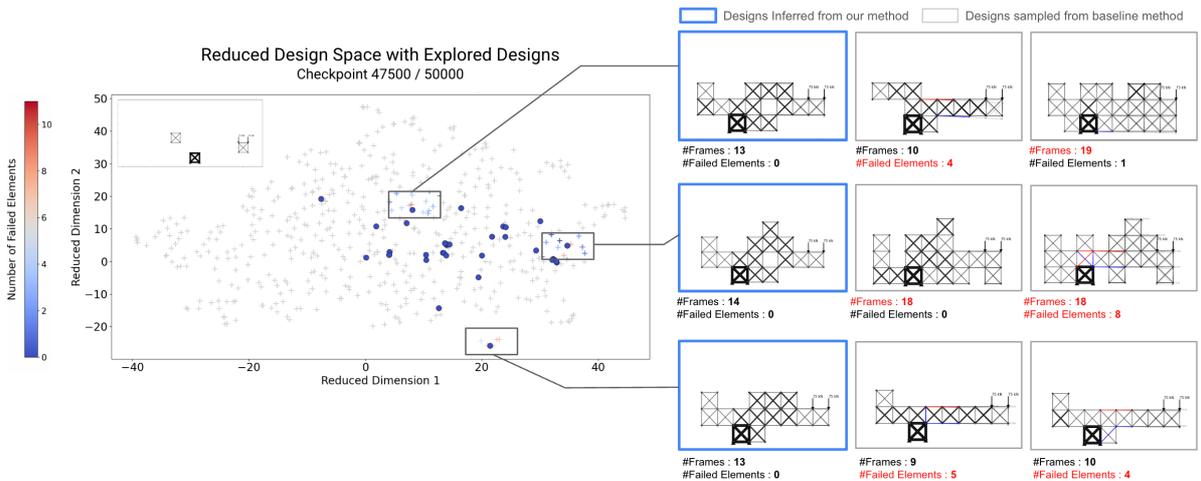

Figure 8: The trained policy generates multiple designs that satisfy the stiffness requirements and minimize material in contrast to geometrically similar naïve designs that either are not stiff enough or use excessive material.

Finally, as shown in Figure *9*, a comparison of the high performing design instances from the baseline and encountered during the search shows that the proposed method encounters high performing instances approximately 110 times more frequently than the random baseline, verifying its efficiency.

*11*

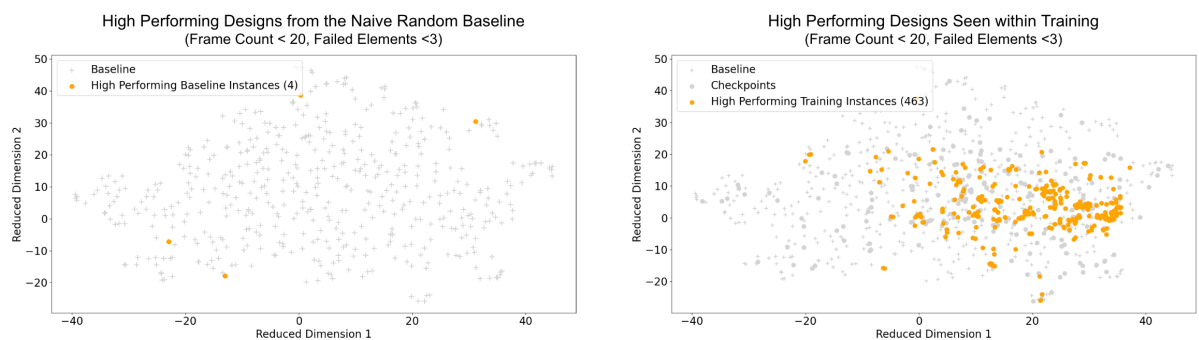

Figure 9: High performing designs (frame count < 20 and failed element count < 3) encountered in the training of the proposed method (463 / 500) compared to the baseline method (4 /500).

## 5. Conclusion

This paper proposes and tests a reinforcement learning framework for performance-driven structural design that combines bottom-up design generation with learned heuristics to efficiently search large combinatorial design spaces. The core novelty of the framework lies in adapting the general logic of reinforcement learning to address the specific complexities of structural design which is possible with the proposed formulation of the discrete structural design problem as a sequential decision-making process, and a human learning inspired phased training algorithm makes possible the learning of the highly non-linear relationship mapping partial geometry to potential structural performance.

The capability of this framework is demonstrated with the challenging task of designing cantilevers using modular truss frames where elegant solutions are not obvious. The trained policy returns distinct, high-performing designs tailored to varying boundary conditions, exhibiting expert-like strategies in overall form of the design and material allocation. Analysis of the search process within the design space during the training of the policy reveals how the proposed method efficiently guides the search across a vast combinatorial design space search towards high performing designs.

These findings highlight the potential to develop computational performance driven design frameworks to go beyond generating high-performing solutions, to emulate expert-like reasoning: learning not only what constitutes a good design, but how it is constructed, and encoding and transferring reasoning for structural design.